\documentclass[aip,jmp,amsmath,amssymb,reprint]{revtex4-1}
\usepackage[T1]{fontenc}
\usepackage[latin9]{inputenc}
\usepackage[letterpaper]{geometry}
\usepackage{textcomp}
\usepackage{amsmath}
\usepackage{graphicx}
\usepackage{amssymb}

\begin{document}

    \title{{\normalsize Influence of disordered edges on transport properties in graphene}}
    \author{{\normalsize D. Smirnov}}
    \email{smirnov@nano.uni-hannover.de}
    \address{Institut f\"ur Festk\"orperphysik, Leibniz Universit\"at Hannover, 30167 Hannover, Germany}

    \author{{\normalsize G. Yu. Vasileva}}
    \address{Institut f\"ur Festk\"orperphysik, Leibniz Universit\"at Hannover, 30167 Hannover, Germany}
    \address{Ioffe Institute, Russian Academy of  Sciences, 194021 St. Petersburg, Russia}

    \author{{\normalsize J. C. Rode}}
    \address{Institut f\"ur Festk\"orperphysik, Leibniz Universit\"at Hannover, 30167 Hannover, Germany}

    \author{{\normalsize C. Belke}}
    \address{Institut f\"ur Festk\"orperphysik, Leibniz Universit\"at Hannover, 30167 Hannover, Germany}

    \author{{\normalsize Yu. B. Vasilyev}}
    \address{Ioffe Institute, Russian Academy of Sciences, 194021 St. Petersburg, Russia}

    \author{{\normalsize Yu. L. Ivanov}}
    \address{Ioffe Institute, Russian Academy of Sciences, 194021 St. Petersburg, Russia}

    \author{{\normalsize R. J. Haug}}
    \address{Institut f\"ur Festk\"orperphysik, Leibniz Universit\"at Hannover, 30167 Hannover, Germany}

    \begin{abstract}
        The influence of plasma etched sample edges on electrical transport and doping is studied. Through electrical transport measurements the overall doping and mobility are analyzed for mono- and bilayer graphene samples. As a result the edge contributes strongly to the overall doping of the samples. Furthermore the edge disorder can be found as the main limiting source of the mobility for narrow samples.
    \end{abstract}

    \maketitle
    Since its discovery in 2004 graphene was praised as a new material with different  possibilities in technical applications\cite{Erscheinungspaper2004[1], Geimreport}. It shows high mobility \cite{Mobility1,Mobility2} on silicon/silicon dioxide substrates on which it can be easily gated. Theoretical mobility limits were calculated \cite{Limitmob} and confirmed in various experimental studies\cite{Overview, Geimreport}. Transferring graphene on better and smoother substrates, e. g. Boron Nitride \cite{BN1}, or removing the substrate completely \cite{Suspended1, Suspended2}, lifted that limit and mobilities of over $1\cdot 10^{6}\,\mathrm{cm^{2}/Vs}$ were measured \cite{Sus1Million}. However these values only refer to non structured samples with undamaged edges. Edge disorder, introduced by various structuring techniques, can further limit the transport properties. Its negative impact is visible in mesoscopic transport measurements in graphene on silicondioxid \cite{SiO2Limits} as well as on Boron Nitride\cite{BNlimits}. Furthermore Raman spectroscopy\cite{Raman1, Raman2, raman2} and chemical doping studies\cite{CarbonDoping} hint towards the edge disorder not only as a mobility limit but a further doping source. Investigations of different kind of edge disorder were performed in previous studies on graphene nanoribbons (GNR) \cite{Z1}. It was shown that plasma etched or similar prepared GNR exhibit disordered sample etches and can lead to a reduction of conductance hinting to a further scattering mechanism \cite{Kats}. Furthermore the influence of such disorder on electrical transport was studied on samples with varying width\cite{W1,W3,W2} and an effect on transport properties was observed. Such disorder was also confirmed through Raman spectroscopy studies of GNRs edges \cite{raman1}. However a quantified evaluation of these edge effects were not presented, yet.

    In this letter a quantified study is performed on mono- and bilayer samples. All flakes were structured in a similar Hall bar geometry with areas of different width. That specific shape allows to investigate the edge doping as well as the influence of edge disorder on the electrical transport in samples with equal bulk doping. Several graphene flakes were investigated within the scope of this study, showing similar results. In this letter, a monolayer and a bilayer sample are presented.
      \begin{figure}[t!]
            \centering
            \includegraphics{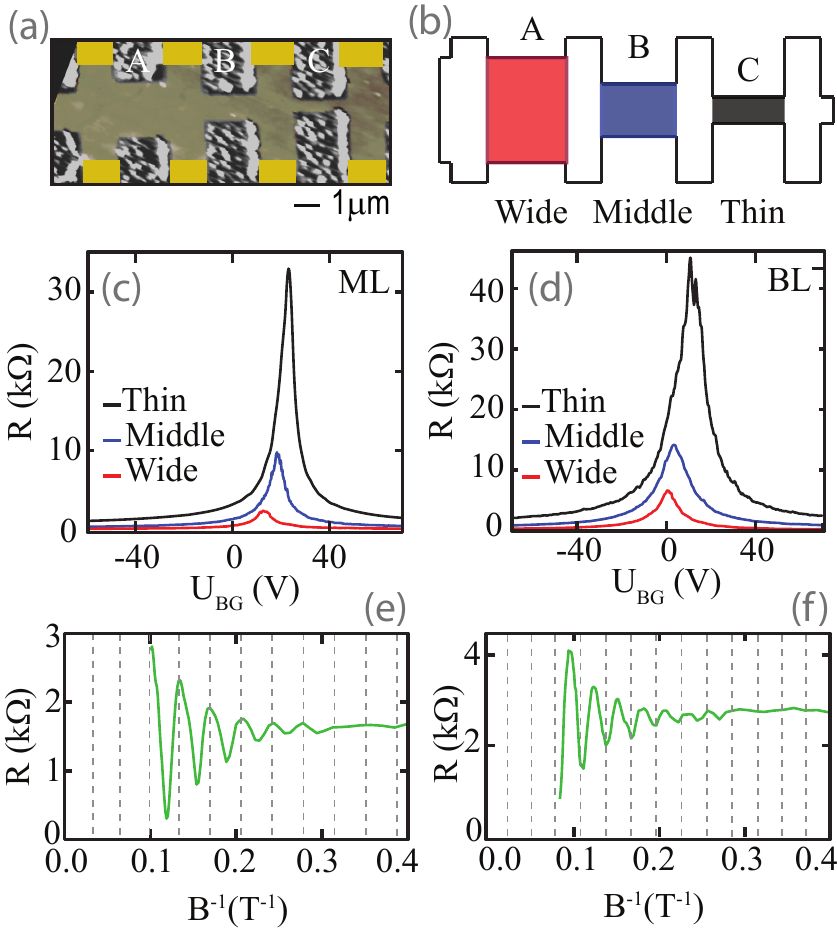}
            \caption {(a) An atomic force microscope picture showing the monolayer sample using false colors. The contact positions are represented through yellow areas. (b) a schematic  picture of the Hall bar sample geometry. Four probe resistance measurements over the different width areas for mono- (c) and bilayer (d) sample. (f) and (g) show the Shubnikov- de Haas measurements versus the inverse magnetic field for mono- and bilayer, respectively. }
            \label{Fig1}
        \end{figure}

    The sample preparation was done as following: Graphene flakes were placed on a silicon/ silicondioxide substrate. Afterwards the number of layers was analyzed using optical microscopy \cite{MakingGrapheneVisible}. Plasma Oxygen Etching was used to edge the flakes into the geometrical shape, shown in Fig. \ref{Fig1}(b). It is a Hall bar, which is divided into three different regions, named "wide", "middle", and "thin". Each part has the same length but differs in width. The length of each area is $2.4\,\mathrm{\mu m}$ for the monolayer sample and $1.8\,\mathrm{\mu m}$ for the bilayer, respectively. The Hall bar width is different for every area: $2\,\mathrm{\mu m}$ (wide), $1\,\mathrm{\mu m}$ (middle), and $0.6\,\mathrm{\mu m}$ (thin) for the monolayer and $1.8\,\mathrm{\mu m}$, $1\,\mathrm{\mu m}$, and $0.5\,\mathrm{\mu m}$ for the bilayer sample. Figure \ref{Fig1}(a) shows an Atomic Force Microscope (AFM) image of a monolayer device. To reduce the overall doping the samples were mechanically cleaned by the AFM in contact mode\cite{AFMCleaning}. Afterwards chromium/ gold contacts were evaporated. After the preparation process, the samples were loaded into a $^{4}\mathrm{He}$ evaporation cryostat and measured at a base temperature of $1.5\,\mathrm{K}$ and a perpendicular magnetic field up to $13\,\mathrm{T}$. The resistance was measured with a lock-in amplifier using an AC current of $100\,\mathrm{nA}$ with a frequency of $17.777\,\mathrm{Hz}$.
  
    The characterization of the samples was conducted for each area using a four terminal set-up. Magnetotransport measurements were performed to confirm the number of layers. Figure \ref{Fig1}(e) and (f) show the longitudinal resistance versus the inverse magnetic field. Shubnikov-de Haas oscillations are visible and the Berry phase can be extracted through extrapolation to zero. The Berry Phase is $\pi$ for the monolayer as expected and $2\pi$ for the bilayer sample, which confirms the previous contrast analysis. Figure \ref{Fig1}(c) and (d) show the resistance measurements versus the backgate voltage $U_{BG}$ for different areas and samples. For each area a field effect is observed. However the position and the resistance of the charge neutrality point differs. In both cases the mono- and the bilayer sample follow the same trend. The thin part of both samples exhibits the highest resistance reaching $33\,\mathrm{k\Omega}$ for the monolayer sample and $45\,\mathrm{k\Omega}$ for the bilayer sample at the charge neutrality point. The resistance decreases with increasing width as expected for the geometry. In contrast, the dependence of the charge neutrality point position is not that simple. The overall doping of both samples is in the positive backgate voltage region, meaning that both samples are p-doped. However the doping concentration differs for every area. Because of the AFM cleaning the bulk of both samples can be ruled out as the origin of the doping change. However a disordered edge can act as a p-doping source, as was shown in previous works \cite{CarbonDoping}. To analyze the amount of edge doping in these samples a simple approach is proposed. The overall doping amount of the sample $N_O=n_O A$, is the sum of the bulk doping $N_B=n_B A$ and the edge doping $N_E=n_E L$,
    \begin{equation*}
        N_O=N_B+N_E,
    \end{equation*}
    with $n_O, n_B, n_E$ being the doping concentration and $A, L, W$ the area, length, and width of the doped region. One can rewrite that approach to the doping concentration, leading to a width dependence:
    \begin{equation*}
        n_O=n_B+n_E\cdot \frac{1}{W}.
    \end{equation*}

   From the backgate voltage at the charge neutrality point the overall doping can be calculated for all areas and both samples.

    \begin{figure}[t!]
        \includegraphics{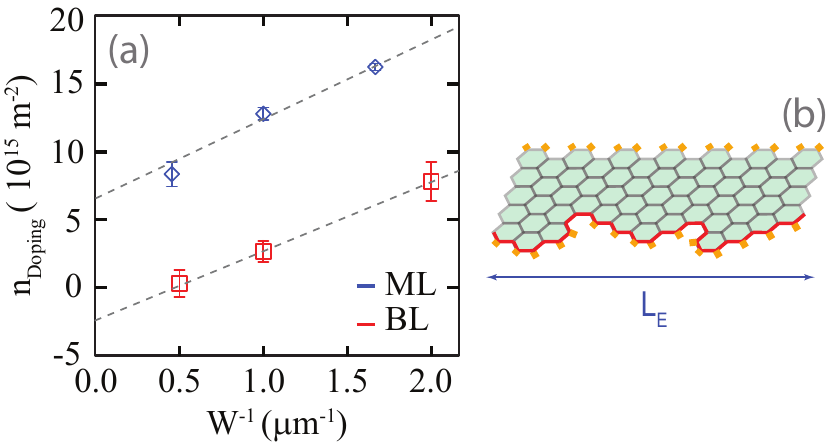}
        \caption{(a) Dependence of the overall doping on sample width. The overall doping was plotted versus the inverse area width for the mono- (blue) and bilayer (red) sample. The shown data has a linear dependence which is clarified with a dashed line. (b) A schematic picture of a sample with a perfect and a disordered edge to clarify their doping contribution.  }
        \label{Fig2}
    \end{figure}

    Figure \ref{Fig2}(a) shows the overall doping concentration plotted versus the inverse Hall bar width. For both samples a linear dependence of the doping concentration is observed. Using the bulk-edge-doping approach shown above, the individual  doping contributions components can be calculated for both samples. The bulk doping concentration is $p_B=6.6\cdot 10^{15}\,\mathrm{m^{-2}}$ and $n_B=2.3\cdot 10^{15}\,\mathrm{m^{-2}}$ for the mono- and bilayer, respectively. The edge doping concentration is $p_E=6.5\cdot 10^{9}\,\mathrm{m^{-1}}$ for the monolayer and an almost equal concentration of $5.1\cdot 10^{9}\,\mathrm{m^{-1}}$ for the bilayer sample leading to effective 2D-doping at the edge. To demonstrate the contribution of the doping components the actual doping amount is calculated for the thinnest area for the bilayer sample, where electrons $P_B\approx 2100$ and holes $N_E\approx 9000$ are introduced into the system through the bulk and the edge. One can clearly see the domination of the edge doping in this area. It is higher than the bulk contribution by a factor of $\approx 4.5$, introducing holes as a main doping type. Such charge localization at sample edges was observed in ultra thin graphene nanoribbons \cite{Z1}. Furthermore it was found that additional edges in form of a cut or defects in the graphene bulk constitute a p-doping source and can be used to create p-n junctions \cite{AFMPN}.

    It is clear that the edge of the sample can be seen as a doping source. Furthermore we can assume that the adatoms causing the edge p-doping are oxygen compounds that adjusted itself on the graphene edge during the plasma oxygen structuring process, as is shown in a schematic in Fig.~\ref{Fig2}(b) (orange lines). To calculate the efficiency for edge doping, it is first assumed that a dangling bond can contribute to the overall doping by a count of $1$ doping carrier. However, assuming a zigzag edge, it is only possible once per $0.246\,\mathrm{nm}$, as is shown in Fig.~\ref{Fig2}(b). Taking into account that every side of the sample contributes to the edge doping an approximate efficiency of $0.8$ is calculated for the monolayer. As one can extract from the slopes in Fig.~\ref{Fig2}(a), the bilayer exhibits almost the same edge doping contribution and doping efficiency as the monolayer sample. Furthermore the doping efficiency of the plasma etched edges is comparable to an intentional chemical edge doping with hydrogen silsesquioxane, which was investigated in Ref. \onlinecite{CarbonDoping}. The resulting efficiency is only slightly higher (0.85) than of the observed values in the presented study.

\begin{figure}[t!]
        \includegraphics{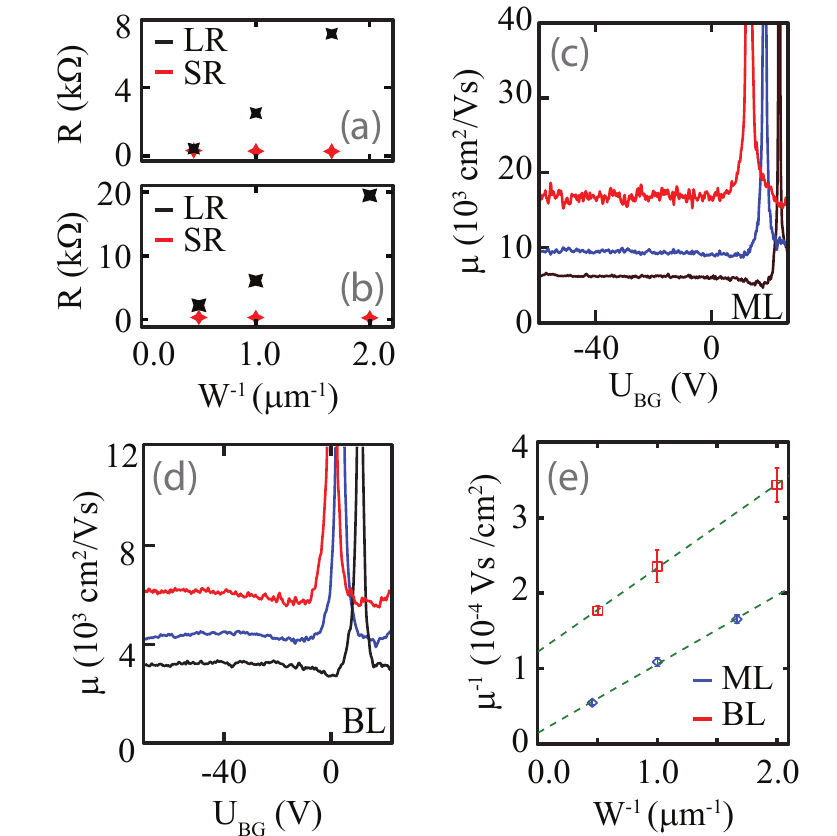}
        \caption{Analysis of the effect of the edge disorder on the electrical transport: (a) and (b) show the long (black) and short (red) range resistance for a fixed charge carrier concentration of $p=6.53\cdot 10^{15}\,\mathrm{1/cm^{2}}$ for the mono- and bilayer, respectively. (c) and (d) show the mobility plotted versus the backgate voltage for the monolayer and bilayer, respectively. (e) shows the dependence of the inverse mobility of different areas on the inverse width. The linear behavior is presented through dashed lines. }
        \label{Fig3}
    \end{figure}

    We further analyze the effect of the disordered plasma etched graphene edge on the electrical transport. The mobility was calculated from the measured resistance versus the backgate-voltage shown in Fig.~\ref{Fig1}(c) and (d). For this analysis the resistance was split up into two different parts: $R_{SR}$ component caused by short range resistance and $R_{LR}$ caused by long range resistance. Both components can be separated using the constant mobility Ansatz \cite{CMA}. In contrast to the doping the short range resistance component stays almost constant throughout all sample areas: $290\,\mathrm{\Omega}$ (wide), $260\,\mathrm{\Omega}$ (mid), and $240\,\mathrm{\Omega}$ (narrow) for the monolayer and $340\,\mathrm{\Omega}$ (wide), $320\,\mathrm{\Omega}$ (mid), $275\,\mathrm{\Omega}$ (narrow) for the bilayer sample, respectively. Figure~\ref{Fig3}(a) and (b) shows the comparison between the short and the long range resistance component for a fixed charge carrier concentration for mono- and bilayer sample, respectively. As one can see the short range component undergoes a slight change of  $\approx 50\,\mathrm{\Omega}$ and is tiny compared to the change in the long range resistance, which is in the order of $\mathrm{k\Omega}$. Furthermore the difference of the short range component with changing width is quite small in comparison to the total short range resistance amount. The origin of the short range component can be located in the bulk and the edge of the sample. However the fact that the short range resistance stays almost constant while the width and with that the area changes indicates that the main short range scattering contributors are located at the edge of the investigated samples, which has the same length for all the samples.

     Subsequently, the constant mobility is calculated from long range resistance component, i. e. the overall resistance after subtracting the short range component. The results are shown in Fig.~\ref{Fig3}(c) and (d). For both samples the mobilities differ with changing width. Additionally the monolayer exhibits a significantly higher mobility for every region as the bilayer, which is an expected behavior for increasing number of layers \cite{MobLayers}. Interestingly, an overall dependence similar to the doping analysis can be observed: The highest mobility is obtained for the wide region (monolayer: $18000\,\mathrm{cm^{2}/Vs}$, bilayer: $5500\,\mathrm{cm^{2}/Vs}$) and the lowest for the narrow region (monolayer: $6000\,\mathrm{cm^{2}/Vs}$, bilayer: $2600\,\mathrm{cm^{2}/Vs}$). The middle section exhibits an intermediate mobility $10000\,\mathrm{cm^{2}/Vs}$ for mono- and $4600\,\mathrm{cm^{2}/Vs}$ for bilayer. Furthermore a significant difference between holes and electrons was not observed.
    
     Hence it follows the same trend as the overall doping, the inverse mobility is plotted versus the inverse width, which is shown in Fig. \ref{Fig3}(e). Equivalent to the overall doping concentration it is showing a linear increase with inverse width in the measurement range. The dependence of the mobility on the width is hinting towards an additional scattering mechanism at the sample edges. Since this correlation is detected in the mobility calculated from the long range resistance component, the origin of this mechanism is most likely caused by the additional edge doping\cite{DAcc} shown above.

    By fitting the data in Fig. \ref{Fig3}(e) linearly and extrapolating to an infinite sample width, the bulk mobility component can be estimated to $0.75\pm 0.05\,\mathrm{m^{2}/Vs}$ for the bilayer and $5.5\pm 1\,\mathrm{m^{2}/Vs}$ for the monolayer sample. Both bulk mobilities exceed the measured mobilities proving the limiting nature of the edge disorder. However it is important to notice, that these same results are an extrapolation of the observed mobilities and not measured. 
    
     As stated above, previous analysis were performed on the topic of edge disorder leading to an increasing resistance \cite{W1,W3,W2}. Our sample geometry allowed to investigate the edge disorder systematically excluding other effects. By extracting the electronic properties from the field effect for every different region we were able to analyze the short and the long range component independently. Our results show that the edge affects both, however, the long range far more than the short range resistance, subsequently influencing the overall mobility greatly. Therefore the edge doping can be seen as a strong scattering mechanism determining the electrical properties even in large, $ \mathrm{\mu m}$-sized samples.

    In conclusion, we have reported an investigation of the influence of the edge disorder on the electrical transport in mono- and bilayer graphene. Our devices allowed to investigate the dependency of various properties on the width of each single sample. We showed how the edge influences the electrical transport greatly, dominates the overall doping, and acts as an additional scattering mechanism.
    \newline
    $\,$ \newline
    We acknowledge the financial support by the DFG via SPP 1459 and the Russian Foundation for Basic Research (Grant
    No. 13-02-00326 a). We are grateful to A. P. Dmitriev, for helpful discussions. Authors D.~S. and G.~Yu.~V. contributed equally to this work.

\end{document}